\documentclass[aps,prb,preprint,groupedaddress,showpacs,]{revtex4}
\usepackage{graphicx}
\usepackage{helvet}
\usepackage{amsmath,amssymb}
\usepackage{bm}

\makeatletter
\newcommand{\Epol}{$E_{\rm pol}$}
\newcommand{\LT}{$L_{T}$}
\newcommand{\LV}{$L_{V}$}
\newcommand{\Vnorm}{$\bar{V}$}

\newcommand{\muzero}{${\mu}_{0}$}
\makeatother
\begin{document}

\title{Thermal Generation of Spin Current in a Multiferroic Helimagnet}
\author{R.\ Takagi,$^{1}$\ Y.\ Tokunaga,$^{1,2}$\ T.\ Ideue,$^{3}$\ Y.\ Taguchi,$^{1}$\ Y.\ Tokura$^{1,3}$,\ and\ S.\ Seki,$^{1,4}$}
\affiliation{$^{1}$RIKEN Center for Emergent Matter Science (CEMS), Wako, 351-0198, Japan \\
$^{2}$Department of Advanced Materials Science, University of Tokyo, Kashiwa, 277-8561, Japan \\
$^{3}$Department of Applied Physics and Quantum Phase Electronics Center (QPEC), University of Tokyo, Tokyo, 113-8656, Japan \\
$^{4}$PRESTO, Japan Sience and Technology Agency (JST), Tokyo, 102-8666, Japan}
 
\date{\today}
\pacs{75.30.Ds, 75.70.-i}

\begin{abstract}
 We report the experimental observation of longitudinal spin Seebeck effect in a multiferroic helimagnet Ba$_{0.5}$Sr$_{1.5}$Zn$_{2}$Fe$_{12}$O$_{22}$. Temperature gradient applied normal to Ba$_{0.5}$Sr$_{1.5}$Zn$_{2}$Fe$_{12}$O$_{22}$/Pt interface generates inverse spin Hall voltage of spin current origin in Pt, whose magnitude was found to be proportional to bulk magnetization of Ba$_{0.5}$Sr$_{1.5}$Zn$_{2}$Fe$_{12}$O$_{22}$ even through the successive magnetic transitions among various helimagnetic and ferrimagnetic phases. This finding demonstrates that the helimagnetic spin wave can be an effective carrier of spin current. By controlling the population ratio of spin-helicity domains characterized by clockwise/counter-clockwise manner of spin rotation with use of poling electric field in the ferroelectric helimagnetic phase, we found that spin-helicity domain distribution does not affect the magnitude of spin current injected into Pt. The results suggest that the spin-wave spin current is rather robust against the spin-helicity domain wall, unlike the case with the conventional ferromagnetic domain wall.
\end{abstract}

\maketitle

 Recently, the concept of spin current, i.e. the flow of spin angular momentum or magnetic moment, attracts much attention as a potential alternative to the charge current with better energy efficiency.\cite{Maekawa,Sharma,Nagaosa,BauerReview}
 Spin current is generally carried by spin-polarized conduction electrons in metals or spin waves (i.e. propagating spin precession) in insulators, the latter of which have many advantages for the dissipationless control, such as the long propagation length and absence of Joule heat loss coming from the simultaneous flow of charge current.\cite{Kajiwara} Spin Seebeck effect is a phenomenon in which spin current is generated by the simultaneous application of temperature gradient (${\nabla}T$) and magnetic field ($H$) to magnetic materials.\cite{SSI,LSSE_APL,KikkawaANE,SSEins,SSEsemicon} When a paramagnetic metal (PM) layer is attached to the magnetic insulator, the thermally-induced spin-wave spin current is injected into the PM layer through the interfacial spin-exchange interaction. The injected spin current ($\vec{J_{S}}$ flowing normal to the interface) transforms into an electric voltage in PM due to the inverse spin Hall effect (ISHE),\cite{LSSE_APL,KikkawaANE,YIGspinpomping} which originates from the spin-orbit interaction and follows the relation
\begin{equation}\label{ISHE}
{\vec{V}_{\rm ISHE}}\ {\propto}\ {L_{V}}({\vec{J_{S}}} {\times} {\vec{\sigma}}).
\end{equation}
Here, ${\vec{\sigma}}(|| H)$ is carried spin angular momentum and $L_{V}$ is the gap distance between the electrodes for the voltage measurement. Previous studies of thermally-induced spin-wave spin current have mostly focused on a limited number of ferromagnets or ferrimagnets such as rare-earth iron garnet $R_{3}$Fe$_{5}$O$_{12}$ (including Y$_{3}$Fe$_{5}$O$_{12}$ (YIG)), spinel ferrite $M$Fe$_{2}$O$_{4}$ ($M$: $3d$ transition metal), and hexaferrite BaFe$_{12}$O$_{19}$.\cite{YIGspinpomping,SpinCurrentFerrite} More recently, the emergence of spin Seebeck effect has been reported for different classes of magnetic insulators such as paramagnets with short-range spin correlation (Gd$_{3}$Ga$_{5}$O$_{12}$ and DyScO$_{3}$)\cite{PMSSE} as well as collinear uniaxial antiferromagnets (Cr$_{2}$O$_{3}$ and MnF$_{2}$),\cite{Cr2O3,MnF2} the latter of which suggests that the antiferromagnetic spin wave can also be a carrier of spin current. However, the localized magnetic moments often order in a more complex manner than simply antiferromagnetic as exemplified by the helical or long-wavelength spin textures in frustrated magnets. It is an important issue to clarify the universal relationship between spin current and spin wave in such noncollinear spin textures.  

 In this Letter, we report the experimental observation of spin Seebeck effect in a multiferroic helimagnetic insulator Ba$_{0.5}$Sr$_{1.5}$Zn$_{2}$Fe$_{12}$O$_{22}$ (BSZFO). Temperature gradient applied normal to the BSZFO/Pt interface generates the inverse spin Hall voltage of spin current origin in Pt, whose magnitude turns out to be proportional to bulk magnetization of BSZFO even under the successive magnetic transitions among various helimagnetic and ferrimagnetic phases. This finding demonstrates that the helimagnetic spin wave can also be an effective carrier of spin current. It is known that BSZFO hosts spin-driven ferroelectricity in an $H$-induced helimagnetic phase,\cite{Kimura,KimuraReview,TokuraReview} where the sign of electric polarization $P$ is coupled with the clockwise (CW)/counter-clockwise (CCW) manner of spin helical sense (i.e. spin helicity). In this ferroelectric helimagnetic state, we found that electric-field ($E$)-induced modification of CW/CCW spin-helicity domain distribution does not affect the magnitude of spin current injected into Pt. The above results suggest that the spin-wave spin current is rather robust against the spin-helicity domain wall, unlike the case with the conventional ferromagnetic domain wall. 

 The target material Ba$_{0.5}$Sr$_{1.5}$Zn$_{2}$Fe$_{12}$O$_{22}$ is a member of Y-type hexaferrites.\cite{Momozawa_MagStructure,Momozawa_phasediagram} The crystal structure belongs to the trigonal space group $R{\bar{3}}m$ and is composed of alternating stacks of $S$ and $L$ blocks along the hexagonal $c$ axis as displayed in Fig. 1(a). In these blocks, the magnetic moments on Fe sites collinearly lie within the $ab$ plane due to in-plane anisotropy. BSZFO exhibits a helimagnetic order below 326 K due to magnetic frustration.\cite{momozawa} When magnetic field is applied within the $ab$ plane, the system undergoes successive magnetic transitions as represented by $H$-$T$ magnetic phase diagram in Fig. 1(b). Figure 1(c) shows the proposed magnetic structure for each magnetic phase.\cite{Momozawa_phasediagram} The magnetic phase is characterized by the magnetic modulation vector ${\vec{q}}\ {\sim}\ (0, 0, 3{\delta})$, and the screw-like helical spin state is realized with incommensurate spin modulation period $0 < {\delta} < 1/2$ under zero magnetic field. As the magnitude of in-plane $H$ increases, the system goes through intermediate I1, I2, and I3 phases, and finally the collinear ferrimagnetic state is stabilized. Here, the I1, I2, and I3 phases host noncollinear spin arrangement with commensurate magnetic modulation period ${\delta} = 1/4$, $1/2$, and $1/2$, respectively. Remarkably, the I3 phase was found to show a magnetically-induced electric polarization.\cite{Kimura,KimuraReview,TokuraReview} While a fan-like spin structure has originally been proposed for this I3 phase,\cite{momozawa} it is inconsistent with the observed inversion symmetry breaking and later the canted screw (or transverse conical) spin texture as shown in Fig. 4(d) has been proposed for the ferroelectric phase by the analogy with other Y-type hexaferrite compounds such as Ba$_{2}$Mg$_{2}$Fe$_{12}$O$_{22}$.\cite{Anneal,KimuraReview,Ishiwata} A similar ferroelectric phase was also observed in several other Y-type hexaferrites such as Ba$_{2}$(Mg$_{1-x}$Zn$_{x}$)$_{2}$Fe$_{12}$O$_{22}$ and Ba$_{0.5}$Sr$_{1.5}$Co$_{2}$Fe$_{12}$O$_{22}$.\cite{BMZFO,BSCFO}
 Here, the microscopic origin of  magnetically-induced ferroelectric polarization has been explained by the spin-current model or the inverse Dzyaloshinskii-Moriya model.\cite{KNB} This model describes the relationship between an local electric dipole moment (${\vec{p}_{ij}}$) and canted spin moments (${\vec{S}_{i}}$ and ${\vec{S}_{j}}$) on the neighboring $i$ and $j$ sites as
\begin{equation}\label{spincurrent}
\vec{p}_{ij}\ {\propto}\ \vec{e}_{ij} {\times} (\vec{S}_{i} {\times} \vec{S}_{j}),
\end{equation}
where ${\vec{e}_{ij}}$ denotes the unit vector between the two sites. $({\vec{S}_{i}} {\times} {\vec{S}_{j}})$ corresponds to the vector normal to the spin-spiral plane, and favors to orient along the $H$-direction. While this model predicts $P = 0$ for the screw spin state with the  $(\vec{S}_{i} {\times} \vec{S}_{j})\ {\|}\ {\vec{q}}$ relationship under $H = 0$, the canted screw spin structure in the I3 phase allows the emergence of $P$ along the direction normal to both ${\vec{q}}$ and $(\vec{S}_{i} {\times} \vec{S}_{j})$. The sign of $(\vec{S}_{i} {\times} \vec{S}_{j})$ represents the clockwise or counter-clockwise manner of spin rotation, and is coupled with the sign of electric polarization (Fig. 4(d) and (e)).\cite{Yamasaki} Without the electric field poling, these two spin-helicity domains (i.e. ${\pm}P$ ferroelectric domains) are degenerate and expected to coexist with equal population ratio.

 Single crystals of BSZFO were grown by a laser floating zone method, and annealed in 10 atm O$_{2}$. They were cut into a rectangular shape with widest faces perpendicular to the crystallographic $[1{\bar{1}}0]$ direction, and polished with diamond slurry and colloidal silica. Figure 1(d) shows a schematic illustration of the measurement setup to detect thermally-induced spin current (the setup A). The sample consists of a bulk BSZFO plate with a thin film of Pt (10 nm) deposited on to the polished $(1{\bar{1}}0)$ surface. To enable the electric-field poling of the BSZFO sample, silver paste was painted on the opposite surface as the backgate. The temperature gradient ${\nabla}T$ is applied normal to the BSZFO/Pt interface, which corresponds to the geometry of longitudinal spin Seebeck effect (LSSE). For this purpose, the sample is sandwiched with two Cu blocks (covered with a thin Al$_{2}$O$_{3}$ film to guarantee good thermal contact and electrical insulation) and their temperatures were stabilized at $T$ and $T$-${\Delta}T$ by using a resistive heater and Cernox thermometer with a temperature controller (Lakeshore 335). We define ${\nabla}T = {\Delta}T/L_{T}$, with $L_{T}$ representing the sample thickness along the temperature gradient direction. In order to evaluate the magnitude of thermally-induced spin current based on Eq. (1), we measured electric voltage $V_{\rm raw}$ in the Pt layer normal to an external magnetic field $H$ applied along the [110] direction. We assumed $V_{\rm raw}$ for ${\Delta}T = 0$ as a background and obtained $H$-odd component of voltage $V$ by $V(H,{\Delta}T)= [(V_{\rm raw}(H,{\Delta}T)-V_{\rm raw}(H,0))-(V_{\rm raw}(-H,{\Delta}T)-V_{\rm raw}(-H,0))]/2$. Magnetization $M$ was measured using a SQUID magnetometer (Magnetic Property Measurement System, Quantum Design). To obtain electric polarization $P$, we measured pyroelectric current with $H$-sweep at a rate of 0.01 T/s by using an electrometer (model 6517A, Keithley) and a superconducting magnet (Physical Property Measurement System, Quantum Design), and integrated it with time. 
 
Figures 2 (a) and (b) exhibit the $H$-dependence of $M$ in BSZFO as well as $V$ in Pt for $T$ = 60 K and 260 K with ${\Delta}T$ = 30 K, measured under the setup A (Fig. 1(d)).  Multiple magnetization steps are observed in the $M$-$H$ curves, which are attributed to the successive evolution of magnetic structures as summarized in Figs. 1(b) and (c). Correspondingly, $V$ increases in a stepwise pattern as a function of $H$. The magnitude of $V$ in Pt is found to be always proportional to $M$ in BSZFO, not only in the ferrimagnetic phase but also in the helimagnetic ones at lower $H$. These results suggest that the observed $V$ originates from the spin-wave spin current in BSZFO carrying spin angular momentum ${\vec{\sigma}}\ {\propto}\ {\vec{M}}$, irrespective of the underlying spin texture. The inset of Fig. 2(a) shows the ${\nabla}T$-dependence of $V$ at 3 T. The $V$ turns out to be proportional to ${\nabla}T$, which is consistent with the relationship $J_{S}\ {\propto}\ {\nabla}T$ as reported for several ferromagnetic materials such as YIG.\cite{LSSE_APL}

 In case of the YIG/Pt system, it has been discussed that the static proximity effect at the interface might induce the ferromagnetic moment in Pt ($M_{\rm Pt}$), since Pt is near the Stoner ferromagnetic instability.\cite{Huang,KikkawaANE} When ${\nabla}T$ is applied normal to $M$ in such a ferromagnetic conductor, the anomalous Nernst effect (ANE) can also generate the similar $H$-antisymmetric electric voltage, given by
\begin{equation}\label{ANE}
V_{\rm ANE}\ {\propto}\ {\vec{M}}_{\rm Pt} {\times} {\vec{\nabla}}T. 
\end{equation}
In this case, the $V$-profile obtained in the setup A consists of $V = V_{\rm ISHE} + V_{\rm ANE}$. To check the possible ANE contamination for the $V$-profile in Fig. 2, we attempted to evaluate the magnitude of pure ANE contribution by employing different measurement configuration as shown in Fig. 3(a) (i.e. the setup B). Here, $H(\ {\|}\ {\sigma})$ is normal to the $c$-axis of BSZFO as in case of the setup A, but is also normal to the BSZFO/Pt interface this time. ${\nabla}T$ is applied parallel to the interface, and the voltage component normal to both ${\nabla}T$ and $H$ is measured on Pt. In this setup B, the inverse spin Hall voltage should vanish due to the relationship $J_{S}\ {\|}\ {\sigma}$ following Eq. (1), and thus the pure contribution of ANE can be estimated.\cite{KikkawaANE} Figure 3(b) shows the $H$-dependence of {\Vnorm} for both setup A and B, where {\Vnorm} denotes the voltage normalized with sample dimensions: {\Vnorm}$=V${\LT}/{\LV}. We found that the {\Vnorm} for the Setup B is more than one order of magnitude smaller than that for the Setup A, which confirms that the ANE voltage is negligible in the present system. These results prove that the $H$-antisymmetric voltage signal observed in Setup A is almost entirely due to the ISHE voltage induced by the LSSE.

 The above relationship $V_{\rm ISHE}\ {\propto}\ M$ has also been reported in the previous LSSE experiments for several ferromagnets and simple uniaxial antiferromagnets.\cite{KikkawaANE,Cr2O3}  However, its validity in the frustrated helimagnets characterized by multiple-step metamagnetic transition is rather unexpected, considering that each magnetic phase possesses distinctive magnon dispersion relationship with unique manner of characteristic spin oscillation. The thermal process excites magnon modes with distributions of energy and wavenumber determined by the Bose factor, which may smooth out the difference of magnon dispersion and lead to the simple relationship $V_{\rm ISHE}\ {\propto}\ M$ throughout the entire process of $H$-induced metamagnetic transitions.

 Now, we focus on the LSSE in the I3 magnetic phase with spin-driven ferroelectricity. Here, the CW/CCW rotation manner of helical spin modulation is coupled with the sign of magnetically-induced electric polarization, and these two degenerated spin-helicity domains (i.e. ${\pm}P$ ferroelectric domains) should coexist with equal population ratio without poling electric field {\Epol}. Considering that the magnetically-induced $P$ emerges along the $[1{\bar{1}}0]$ direction (Figs. 4(d) and (e)) based on Eq. (2) for the setup A (Fig. 1(d)), the application of electric field between the Pt layer and Ag backgate located on the opposite surfaces of BSZFO (i.e. {\Epol} ${\|}\ P$) can modify the population ratio of these two spin-helicity domains. To confirm such {\Epol}-induced change of domain distribution, we measured the $[1{\bar{1}}0]$ component of $P$ as a function of the magnitude of $H \ {\|}$ [110] for BSZFO at $T$ = 50 K. Before the measurement, we applied a poling electric field {\Epol} = 425 kV/m in the paraelectric ferrimagnetic phase at {\muzero}$H$ = 3 T, and then entered the ferroelectric I3 phase at {\muzero}$H$ = 1.2 T while keeping {\Epol}. After these procedures, we removed {\Epol} and performed the measurement of $P$. The inset of Fig. 4(b) is the initial $H$-scan profile just after removing {\Epol} at {\muzero}$H$ = 1.2 T. The spontaneous $P$ (${\sim}$40 ${\mu}$C/m$^{2}$) is confirmed to appear only in the I3 phase, in accord with the previous report. \cite{Kimura} Figure 4(b) indicates the $P$-profile for the successive $H$-scan of $+$3 T${\to}-$3 T${\to}+$3 T after the above poling procedure and once entering the paraelectric ferrimagnetic phase at 3 T, where the magnitude of $P$ was reduced to ${\sim}1/4$ of the initial saturation value but with no further decay during repeated $H$-scans. This is a kind of magnetoelectric memory effect as recently reported for several ferroelectric helimagnets.\cite{Taniguchi,Nakajima} We also confirmed that the sign of $P$ is reversed for the opposite sign of {\Epol}, and $P$ vanishes for {\Epol} = 0. These results suggest that the sizable change in the population ratio of the ${\pm}P$ (i.e. CW/CCW spin helicity) domains and the associated domain wall density should occur between the cases with {\Epol} = 0 and {\Epol} = 425 kV/m, even after turning off the applied {\Epol}. Then, we performed the measurement of LSSE after the same poling procedure. Figure 4(c) shows the $H$-dependence of $V$ in Pt at $T$ = 60 K under the setup A; no obvious change in $V$ was found between the cases with and without the poling procedure. This indicates that the magnitude of spin current injected into Pt is not affected by the density of CW/CCW spin-helicity domain wall in BSZFO. Such a conclusion is also supported by the observed $M\ {\propto}\ V$ relationship during the metamagnetic transitions, where the CW/CCW spin-helicity domain wall exists only in the helimagnetic phase and not in the ferrimagnetic phase.

 In general, spin-wave spin current is expressed as the propagating flow of local net magnetization. In case of the bulk YIG/Pt system, the existence of ${\pm}M$ ferromagnetic domain wall has been reported to cause the magnon scattering and significantly suppress the LSSE signal.\cite{LSSE_APL} In contrast, the CW/CCW spin-helicity domains have the same magnetic susceptibility or $H$-induced magnetization, and thus their domain wall can have little influence on the scattering of spin-wave spin current. Here, the ratio between the average domain size and magnon decay length may be also relevant, since the magnons can reach the Pt layer without the domain wall scattering if the former length scale is much longer than the latter one. According to the recent resonant X-ray microdiffraction experiment with circular polarization, the average size of spin-helicity domain for BSZFO is in the order of ${\sim}100 \ {\mu}$m.\cite{resonantXraymicro,softXraymicro} On the other hand, the magnon decay length in insulating materials often reaches several centimeters as reported for YIG,\cite{Kajiwara} which implies that the spin-wave spin current interacting with the Pt layer indeed passes through the spin-helicity domain walls in the present compound. To fully testify this scenario, further experiment associated with coherent magnon transport using magnetic resonance technique would be useful.
 
 In summary, we have experimentally observed longitudinal spin Seebeck effect in a multiferroic helimagnet Ba$_{0.5}$Sr$_{1.5}$Zn$_{2}$Fe$_{12}$O$_{22}$. Temperature gradient applied normal to the BSZFO/Pt interface generates the inverse spin Hall voltage of spin current origin in Pt, whose magnitude was found to be proportional to magnetization of Ba$_{0.5}$Sr$_{1.5}$Zn$_{2}$Fe$_{12}$O$_{22}$ irrespective of underlying spin textures and even through the multiple-step metamagnetic transitions. This demonstrates that the helimagnetic spin wave can be an effective carrier of spin current. In the ferroelectric helimagnetic phase, $E$-induced modulation of spin-helicity domain distribution does not affect the magnitude of spin current injected into Pt. The above results suggest that the spin-wave spin current is robust against the spin-helicity domain wall, unlike the case with the conventional ${\pm}M$ ferromagnetic domain wall.
 
 The authors thank M. Nakamura, M. Kawasaki, Y. Kaneko and A. Kikkawa for experimental helps. This work was partly supported by the Mitsubishi Foundation, and Grants-in-Aid for Scientific Research (Grants No. 26610109 and No. 15H05458) from the MEXT of Japan and Japan Society for the Promotion of Science (JSPS).

\begin{figure}
\begin{center}
\includegraphics[width=8.5cm,keepaspectratio]{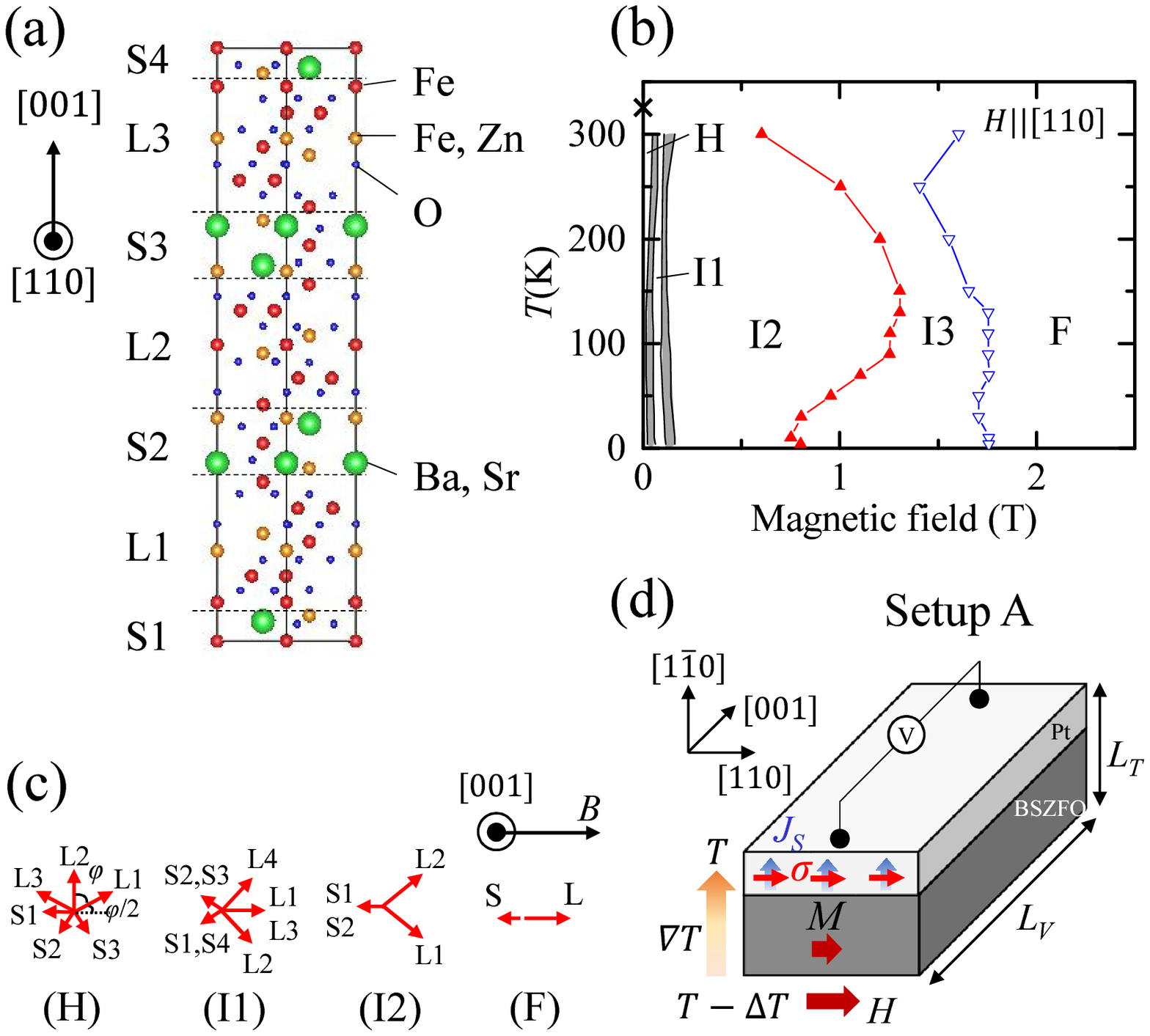}
\end{center}
\caption{(Color online)(a) Crystal structure of Ba$_{0.5}$Sr$_{1.5}$Zn$_{2}$Fe$_{12}$O$_{22}$ (BSZFO), composed of alternating $L$ and $S$ magnetic blocks stacking along $c$-axis. (b) $H$-$T$ magnetic phase diagram for the BSZFO single crystal used in this study, which was determined by measurements of $M$ under $H\ {\|}\ [110]$. A cross mark at zero magnetic field represents the helical magnetic ordering temperature $T_{\rm N}$ = 326 K. (c) Proposed magnetic structures for the helical (H), intermediate I1, I2, and ferrimagnetic (F) phases. Long and short arrows indicate effective magnetic moments of the $L$ and $S$ blocks, respectively. (d) Schematic illustration of the experimental setup for the measurement of longitudinal spin Seebeck effect for the BSZFO/Pt sample (i.e. setup A). $L_{T}$ (thickness of the BSZFO plate along the temperature gradient direction) and $L_{V}$ (gap distance between electrodes on the Pt layer) are 0.16 mm and 0.64 mm, respectively. Blue and red arrows represent the propagating direction of spin current ($J_{S}$) and carried spin angular momentum (${\sigma}$), respectively.}
\label{Fig1}
\end{figure}

\begin{figure}
\begin{center}
\includegraphics[width=8.5cm,keepaspectratio]{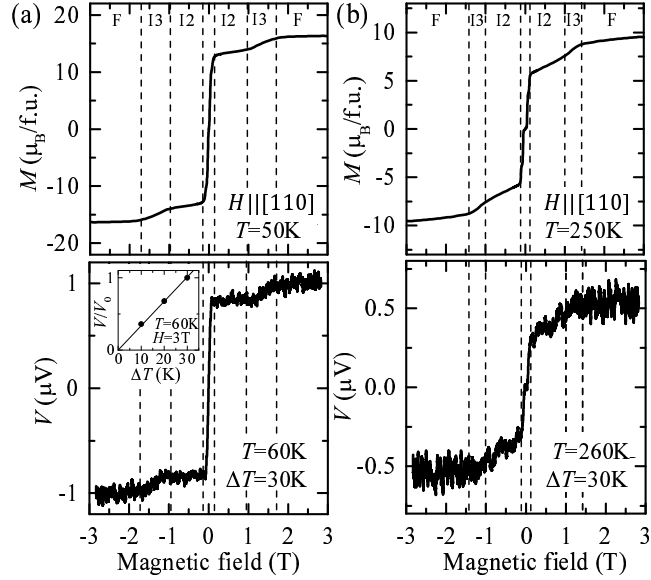}
\end{center}
\caption{(Color online)The upper panels show magnetic field dependence of magnetization for BSZFO with $H\ {\|}\ [110]$ at $T =$ (a) 50 K and (b) 250 K. The lower panels are the corresponding magnetic field dependence of Pt voltage measured for the setup A (Fig. 1(d)) at $T =$(a) 60 K and (b) 260 K with ${\Delta}T=30$  K. The inset of (a) shows the relative magnitude of Pt voltage as a function of ${\Delta}T$ at ${\mu}_{0}H =$ 3 T. Here, $V_{0}$ is defined as the $V$-value obtained with ${\Delta}T$ = 30 K.}
\label{Fig2}
\end{figure}

\begin{figure}
\begin{center}
\includegraphics[width=8.5cm,keepaspectratio]{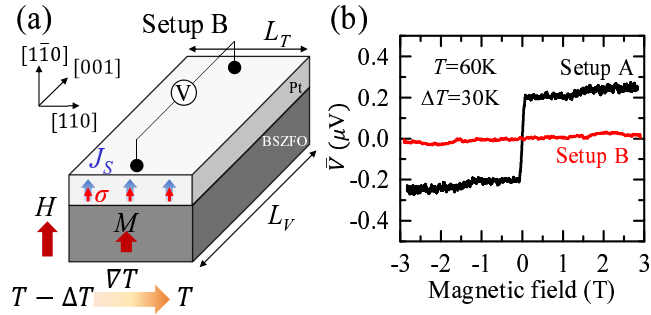}
\end{center}
\caption{(Color online)(a)Schematic illustration of the experimental setup for the measurement of pure anomalous Nernst effect in the BSZFO/Pt sample (i.e. setup B). $L_{T}$ and $L_{V}$ are 0.56 mm and 1.5 mm, respectively. (b) Magnetic field dependence of the normalized Pt voltage ${\bar{V}}$ ($=V L_{T}/L_{V}$) measured for the setup A (Fig. 1(d)) and the setup B.}
\label{Fig3}
\end{figure}

\begin{figure}
\begin{center}
\includegraphics[width=8.5cm,keepaspectratio]{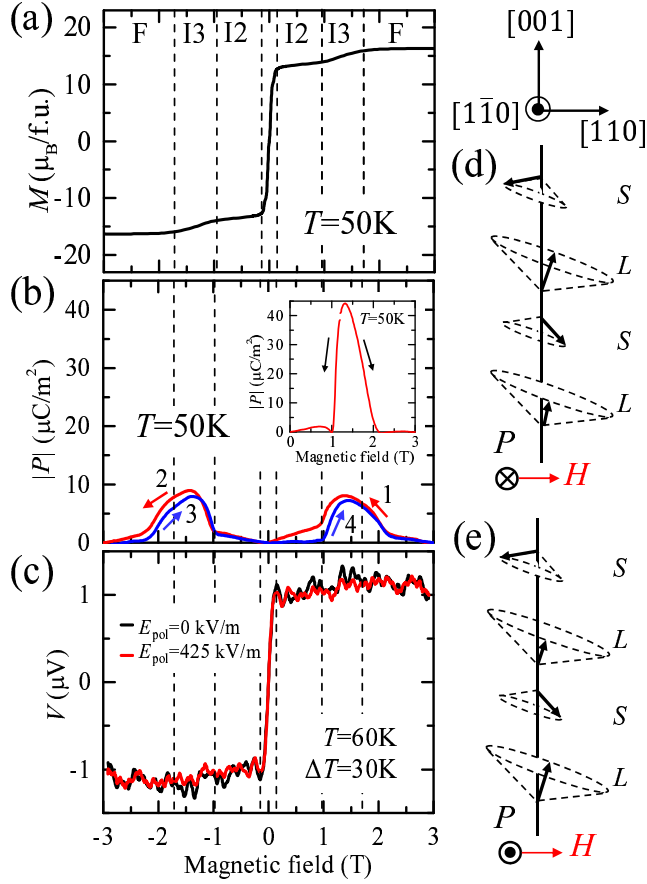}
\end{center}
\caption{(Color online)(a) Magnetization and (b) $[1{\bar{1}}0]$ component of electric polarization as a function of the magnitude of $H\ {\|}\ [110]$ for BSZFO at 50 K. (c) Corresponding magnetic field dependence of Pt voltage measured for the setup A (Fig. 1(d)) at $T = 60$ K and ${\Delta}T=30$ K with and without the electric field poling. To pole the BSZFO crystal, {\Epol} = 425 kV/m was applied at ${\mu}_{0}H = 3$ T, and then we set ${\mu}_{0}H = 1.2$ T. After this poling procedure, we removed {\Epol} and increased the magnetic field up to 3 T, and then started the measurement of the electric polarization in BSZFO or ISHE voltage in Pt with the $H$-sweep of $+3$ T${\to}-$3 T${\to}+$3 T. The initial $P$-profile obtained just after removing {\Epol} at 1.2 T is also plotted in the inset of (b). (d), (e) Canted screw (or transverse conical) spin textures proposed for the I3 magnetic phase. Magnetic field is applied along the [110] direction, and $P$ appears parallel or antiparallel to the $[1{\bar{1}}0]$ direction depending on the clockwise or counter-clockwise manner of helical spin rotation (i.e. spin helicity).}
\label{Fig4}
\end{figure}

\end{document}